# How to Determine the Pion Cloud of the Constituent Quark

S. Baumgärtner[1] and H.J. Pirner
*Institut für Theoretische Physik, Universität Heidelberg, Germany*

K. Königsmann and B. Povh
*MPI für Kernphysik Heidelberg, Germany*

**Abstract**

We calculate the differential cross section for semi-inclusive pion production in electron proton reactions using a model where the physical quark fluctuates with some probability to quark plus pion. The kinematic regions for a determination of this 'pion cloud' are evaluated.

## 1. Introduction

Since Yukawa the nucleon has been perceived as an extended object with a pionic cloud around it. The one-pion exchange interaction between nucleons is a natural consequence of pion emission and absorption by nucleons. The extent of the pionic cloud of the nucleon has, however, been debated. Proponents of the N-N-one-boson exchange model use form factors with cut-off values $\Lambda \approx 2 - 2.5$ GeV, which give a large amount of pion admixture to the nucleon. In an analysis of pion-electroproduction data on the proton [1] an estimate of the pion cloud of the nucleon has yielded a rather small value, namely a 3% admixture of $n\pi^+$ to the naked p state. This small probability would make it unrealistic to use coherent pion emission of the nucleon to create sea quarks in the nucleon, which is a process proposed in Refs. [2].

On the other hand, it is very natural to couple quarks to scalar ($\sigma$) and pseudoscalar mesons ($\vec{\pi}$) in a chirally invariant way. The first coupling leads to the constituent quark mass after spontaneous breaking of the $(\sigma, \vec{\pi})$-symmetry. The second coupling to pions produces a pion cloud of the constituent quark. There is a difference between a pion cloud of the *nucleon* and the pion cloud of the *quark*. Emission and absorption of a pion by the *same* quark determines the pion cloud of the constituent quark. The intermediate states on the nucleon level would be $|3q\pi\rangle$ states of arbitrary excitation energy.

[1]Present address: Alfred Weber- Institut, Grabengasse 14, D-69117 Heidelberg

Allowed intermediate states of the nucleonic pion cloud are $|\, N^*\pi\rangle$ or $|\, \Delta\pi\rangle$ states. In this sense the pion is part of the hadronic system. The analysis of electroproduction at small inclusive mass and correspondingly $z \equiv E_\pi/\nu \rightarrow 1$ will give only the pion cloud of the nucleon, as described in ref. [1]. We suggest that the constituent quark structure is the more interesting quantity to study, as was also proposed in Ref. [3].

The constituent quark picture has been a simple and convincing concept to explain the static properties of baryons and mesons. It is surprising how well magnetic moments and masses of baryons are reproduced. In deep inelastic lepton nucleon scattering the quarks are visible with their momentum distributions and charges. Yet the constituent quark model has never been able to explain the measured structure functions of the proton. Especially the transition with $\mathrm{Q}^2$ from an extended constituent quark with its own internal structure to a pointlike current quark has not been investigated.

At low $\mathrm{Q}^2$, the constituent quark has intrinsic structure, namely a pion and possibly also a gluon cloud. The baryon will be described by a light cone wave function of three constituent quarks. This wave function takes care of the relativistic motion of the quarks and can, in principle, be obtained from a suitable light cone Hamiltonian.

In this article, we want to focus on the phenomenological aspects of the constituent quark. We want to define a procedure to separate its internal structure from the bound state dynamics contained in the quark wave function.

For this purpose, we first analyze in Section 2 the quasi-static properties of the nucleon which are derived from moments of deep inelastic structure functions, namely the Gottfried sum rule, $S_G$, the axial vector coupling, $g_A$, and the spin sum rules, $\Gamma_1^{p,n}$.

These moments depend on integrated pion probability distributions. In the second step (Section 3) we determine the semi-inclusive pion production cross section which gives differential information about the pion cloud of the constituent quark. Section 4 analyzes the experimental possibilities to measure the semi-inclusive process in the relevant $Q^2$, $x_{\mathrm{B}}$, and $z$ range described. Finally we conclude in Section 5.

## 2. The Pion Content of the Constituent Quark

In the naive parton model the proton consists of two up-quarks and one down-quark: $|\, p\,\rangle = |\, uud\,\rangle$. Beside the valence quarks there is a sea of quark antiquark excitations which, however, does not affect the sum rules concerning electric charge, baryon number or strangeness. In an obvious notation, the total content of flavor $i$ in the proton can be written as

$$q_i(x) = q_{i,v}(x) + q_{i,s}(x) + \bar{q}_{i,s}(x) \qquad \text{where } i = \text{u,d,s} \ . \tag{1}$$

Making two assumptions, namely

1. Isospin invariance:

$$u^p(x) \equiv d^n(x) \qquad \text{and} \qquad d^p(x) \equiv u^n(x) \tag{2}$$

2. Flavor symmetry of the sea:

$$u_s(x) \equiv \bar{u}_s(x) \equiv d_s(x) \equiv \bar{d}_s(x) \tag{3}$$

the Gottfried sum rule [4] is derived in this simple model to be

$$S_G = \int_0^1 \frac{\mathrm{d}x}{x} \left[F_2^{ep}(x,Q^2) - F_2^{en}(x,Q^2)\right] = \frac{1}{3} \; . \tag{4}$$

Measurements at CERN [5] have yielded an experimental value of

$$S_G^{exp} = 0.240 \pm 0.016 \tag{5}$$

which implies a substantial violation of the Gottfried sum rule, Eq. 4. This so-called GSR defect can be accommodated in two ways. It has been suggested [7] that contributions from the unmeasured region of small $x$ ($x < 0.004$) could make up this defect. However, recent measurements of the structure functions for very small $x$ (down to $x \sim 10^{-4}$) at HERA [8] show that the NMC extrapolation employed in Eq. 5 is fairly good. The other interpretation, first proposed by Eichten, Hinchliffe and Quigg [9] and assumed in this paper, is that assumption 2 used to derive the Gottfried sum rule is wrong and the sea of light quarks is flavor asymmetric instead, i.e.

$$u_s(x) \equiv \bar{u}_s(x) \neq d_s(x) \equiv \bar{d}_s(x) \; . \tag{6}$$

Now we get a prediction for the Gottfried sum which is different from the value 1/3 depending on flavor asymmetry:

$$S_G = \frac{1}{3} + \frac{2}{3} \int_0^1 \mathrm{d}x \left[\bar{u}_s(x) - \bar{d}_s(x)\right] \; . \tag{7}$$

The observed value of $S_G = 0.240 \pm 0.016$ means that in the sea of the proton there is an excess of d-quarks over u-quarks [10]. The origin of this flavor asymmetry can be understood in the framework of effective chiral quark theory [11,12]. Quarks (that is: massive constituent quarks) can fluctuate into quarks and pions, the only bosonic degrees of freedom considered here. Hence, the physical constituent quark is the coherent sum of a single naked (albeit massive) constituent quark and a state in which the quark is dissociated into constituent quark and pion. It is this idea that we want to term “pion cloud of the constituent quark”.

Before presenting a more detailed discussion in the next section of the properties of this pion cloud of the constituent quark, we want to employ a toy model [9] that already captures the essential features. In this simplified picture a nucleon consists of up- and down-quarks only. In the chiral quark model a u-quark can emit a $\pi^+$ (containing a valence u-quark and a valence $\bar{d}$-antiquark) or a $\pi^0$ as depicted in Fig. 1. The model parameter describing this idea is

$$a = |\, \langle\, d\pi^+ |\, u\, \rangle\, |^{\,2}\ , \tag{8}$$

the probability for an up-quark to turn into a down-quark with the emission of a $\pi^+$. We assume that this fluctuation is small enough to be treated as a perturbation.

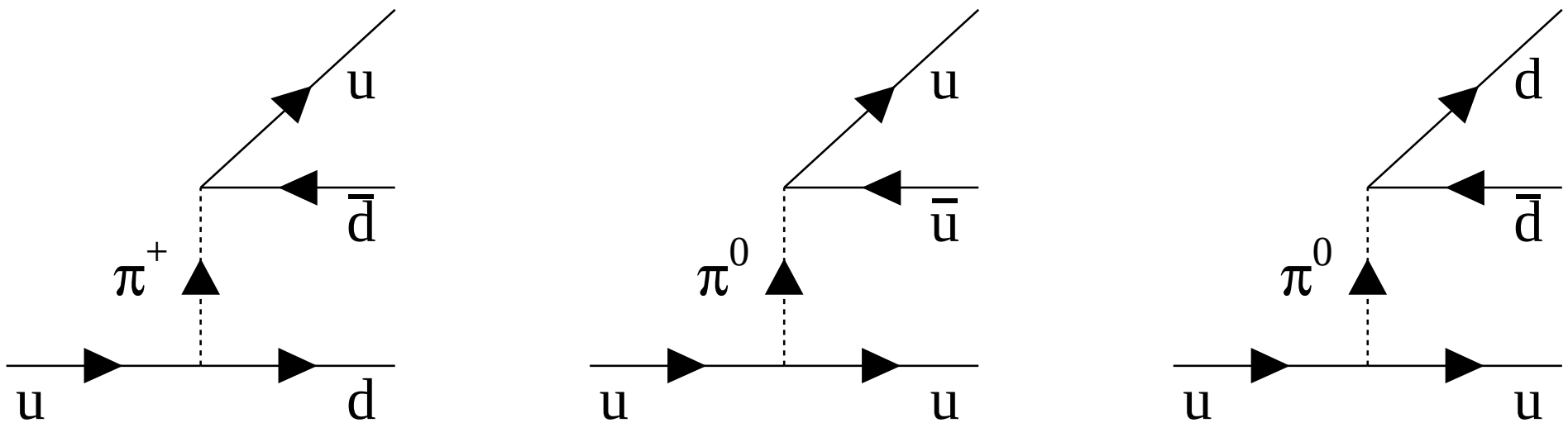

Figure 1. Fluctuation of a u-quark into quark and pion.

In this picture a physical constituent quark ($U$) after one such interaction is a superposition of different (naked) states:

$$\begin{aligned} |\, U\, \rangle &= \sqrt{a}\, |\, \pi^+ d\, \rangle + \sqrt{\frac{a}{2}}\, |\, \pi^0 u\, \rangle + \sqrt{\left(1 - a - \frac{a}{2}\right)}\, |\, u\, \rangle \\ \sum_i e_i^2\, N_{i/U} &= e_u^2 \left[ \left(1 + \frac{1}{4} a\right) N_{u/U} + \frac{a}{4} N_{\bar{u}/U} \right] + e_d^2 \left[ \frac{5a}{4} N_{d/U} + \frac{5a}{4} N_{\bar{d}/U} \right] \end{aligned} \tag{9}$$

and analogously:

$$\begin{aligned} |\, D\, \rangle &= \sqrt{a}\, |\, \pi^- u\, \rangle + \sqrt{\frac{a}{2}}\, |\, \pi^0 d\, \rangle + \sqrt{\left(1 - a - \frac{a}{2}\right)}\, |\, d\, \rangle \\ \sum_i e_i^2\, N_{i/D} &= e_d^2 \left[ \left(1 + \frac{1}{4} a\right) N_{d/D} + \frac{a}{4} N_{\bar{d}/D} \right] + e_u^2 \left[ \frac{5a}{4} N_{u/D} + \frac{5a}{4} N_{\bar{u}/D} \right] \end{aligned} \tag{10}$$

The flavor content of the proton and neutron in this simplified model hence is given by

$$\sum_i e_i^2 N_{i/p} = e_u^2 \left[\left(2+\frac{7a}{4}\right) N_{u/p} + \left(\frac{7a}{4}\right) N_{\bar{u}/p}\right] + e_d^2 \left[\left(1+\frac{11a}{4}\right) N_{d/p} + \left(\frac{11a}{4}\right) N_{\bar{d}/p}\right] \tag{11}$$

and

$$\sum_i e_i^2 N_{i/n} = e_d^2 \left[\left(2+\frac{7a}{4}\right) N_{d/n} + \left(\frac{7a}{4}\right) N_{\bar{d}/n}\right] + e_u^2 \left[\left(1+\frac{11a}{4}\right) N_{u/n} + \left(\frac{11a}{4}\right) N_{\bar{u}/n}\right] \tag{12}$$

One sees: by assuming a pion cloud of the u- and d-constituent quarks (that is: $a \neq 0$) there arises a flavor asymmetry in the nucleon's sea the extent of which depends on the value of the model parameter $a$. Hence, the GSR defect can be expressed in terms of $a$:

$$S_G = \frac{1}{3} - \frac{2a}{3}\,. \tag{13}$$

Of course, the flavor asymmetry represented by the parameter $a$ also affects other static and quasi-static properties of the nucleon. Let

$$\Delta q_i := \int_0^1 \mathrm{d}x \; \left[q_i^{\uparrow}(x) - q_i^{\downarrow}(x)\right] \tag{14}$$

denote the integrated difference of quarks (and antiquarks; see Eq. 1) with helicity parallel ($\uparrow$) and antiparallel ($\downarrow$) to the nucleon's helicity. In the naive quark-parton model with SU(6)-wave functions for a polarized proton one has

$$\begin{aligned} \Delta u &= \frac{4}{3} \\ \Delta d &= -\frac{1}{3} \end{aligned} \tag{15}$$

and gets the following prediction for the axial vector coupling constant $g_A$

$$g_A = \Delta u - \Delta d = \frac{5}{3}\,. \tag{16}$$

If we consider a pion cloud of the constituent quark in the form of the just explained toy model, the spin content of the constituent quark is changed, since the pion is emitted in a $p$-wave state. One gets

$$\begin{aligned} \Delta U &= \frac{4}{3} - \frac{7a}{3} \\ \Delta D &= -\frac{1}{3} - \frac{2a}{3} \\ \Delta S &= 0 \end{aligned} \tag{17}$$

and thus a value of $g_A$

$$g_A = \Delta U - \Delta D = \frac{5}{3}(1-a) \ , \tag{18}$$

which is modified as compared to the naive expectations in Eq. 16 by the existence of a pionic admixture ($a \neq 0$) to the constituent quark.

Similarly, $g_A^0$, defined as the proton matrix element of the flavor singlet axial current

$$\langle\, p \,|\, \bar{\psi}\, \vec{\gamma}\, \gamma_5\, \psi \,|\, p \,\rangle \ =: \ g_A^0\, \chi^\dagger\, \vec{\sigma}\, \chi \ , \tag{19}$$

where $\psi$ = (u,d,s) denotes the quark field and $|\, p\,\rangle$ is the wave function of a proton at rest with Pauli spinor $\chi$, can be expressed using Eq. 17:

$$g_A^0 = \Delta U + \Delta D + \Delta S = 1 - 3a \ . \tag{20}$$

$g_A^0$ measures how much of the proton's helicity is carried by constituent quarks. In the naive quark parton model $a = 0$ and the proton's helicity is carried entirely by quarks ($g_A^0 = 1$). In the model of a pion cloud this value is reduced by the pions' (or sea-quarks') contribution and thus $g_A^0 < 1$.

A rich source of insight into nucleon structure are structure functions for deep inelastic inclusive scattering on polarized nucleon targets. They can also be expressed using quark distribution functions:

$$g_1(x) = \frac{1}{2} \sum_i e_i^2 \left[ q_i^\uparrow(x) - q_i^\downarrow(x) \right] \ . \tag{21}$$

For experimental reasons, instead of the structure functions one rather considers moments of them:

$$\Gamma_1 := \int_0^1 \mathrm{d}x\, g_1(x) \ . \tag{22}$$

In the pion cloud model one obtains for the proton and neutron spin structure functions

$$\Gamma_1^p = \frac{5}{18}(1-2a) \tag{23}$$

$$\Gamma_1^n = -\frac{5}{18}\, a \ . \tag{24}$$

We will now analyse the available experimental data on the above mentioned quantities $S_G$, $g_A$, $g_A^0$ and $\Gamma_1^{p/n}$ to determine the value of $a$.

The Gottfried sum has been measured by the New Muon Collaboration (NMC) at CERN [5]. They determined the ratio $F_2^{\mu n}/F_2^{\mu p}$ in the kinematic range $0.004 < x < 0.8$ and $0.4$ GeV$^2 < Q^2 < 190$ GeV$^2$ and obtained

$$S_G(0.004, 0.8; \langle Q^2 \rangle = 4 \ \mathrm{GeV}^2) = 0.227 \pm 0.006 \tag{25}$$

This value has been extrapolated to $x \to 1$ and $x \to 0$ assuming Regge behavior for small $x$ with the result

$$S_G = 0.240 \pm 0.016 \ . \tag{26}$$

An improved analysis of the data [6] has yielded the value

$$S_G = 0.235 \pm 0.026 \tag{27}$$

which we will employ in our analysis. This improved data analysis also suggested that a so-called shadowing effect occurs when scattering on deuterium targets: the neutron structure function at small $x$ is larger than the difference between the deuterium and the proton structure functions. This effect has not yet been measured directly, but has to be estimated theoretically. Since NMC obtained the neutron structure function as the difference between deuterium and proton structure functions its result for $S_G$ has to be corrected for this shadowing effect. Various calculations on the shadowing effect have already been done [13–15]. For the measured range $x > 0.004$ the authors of [13] obtain a correction to $S_G$ of $\delta^{shadow} S_G = -0.026$. For the whole range $0 < x < 1$ the authors of [14] give a correction between $-0.010$ and $-0.026$. However, comparing these calculations with data for photoproduction shows that the effect is underestimated in [13]. Using a different model for diffractive scattering [15] one obtains a correction to $S_G$ in the measured range $x > 0.004$ of $\delta^{shadow} S_G = -0.043$ and an additional correction in the region not measured ($x < 0.004$) of $\delta^{shadow} S_G = -0.038$. For further analysis we employ the latter result and thus have

$$S_G(0,1;\langle Q^2 \rangle = 4 \ \mathrm{GeV}^2; \mathrm{corrected}) = 0.154 \pm 0.026 \ . \tag{28}$$

Note that the correction for shadowing amounts to 30 % of the original NMC result!

The quantity known with by far the highest precision [16] is the axial vector coupling constant

$$g_A = 1.2573 \pm 0.0028 \ . \tag{29}$$

The spin structure functions $g_1(x)$ of the proton and the neutron have been measured recently by several collaborations. The European Muon Collaboration (EMC) has measured the moment of the proton structure function in the range $Q^2 \geq 10 \ \mathrm{GeV}^2$ (with $\langle Q^2 \rangle = 10.7 \ \mathrm{GeV}^2$) by scattering muons off hydrogen targets [17]:

$$\Gamma_1^p = 0.126 \pm 0.025 \ . \tag{30}$$

At CERN the Spin Muon Collaboration (SMC) [18] has measured the moment of the deuteron structure function in the kinematic range $0.006 < x < 0.6$ and $1 \ \mathrm{GeV}^2 <$

$Q^2 < 30$ GeV$^2$ (with $\langle Q^2 \rangle = 4.6$ GeV$^2$) to be

$$\Gamma_1^d = 0.023 \pm 0.035 \ . \tag{31}$$

At SLAC the E142 experiment has measured [19] the spin structure function of the neutron in the kinematic range $0.03 < x < 0.6$ at $\langle Q^2 \rangle = 2$ GeV$^2$ to be

$$\int\limits_{0.03}^{0.6} \mathrm{d}x \, g_1^n(x) = -0.019 \pm 0.012 \ . \tag{32}$$

From this value an extrapolation yields

$$\Gamma_1^n = \int\limits_0^1 \mathrm{d}x \, g_1^n(x) = -0.022 \pm 0.011 \ . \tag{33}$$

Ellis and Karliner [20] pointed out that extreme care should be taken when comparing these results on polarized scattering because the experiments have been carried out at different $Q^2$. They performed a careful analysis of the experimental results and applied various corrections ($Q^2$-dependence, perturbative QCD corrections, higher twist effects). They find that the data from E142 and SMC are consistent within less than one standard deviation with the Bjorken sum rule ($\int \mathrm{d}x \, (g_1^p(x) - g_1^n(x))$. One sees that the $Q^2$-corrections are essential for understanding the data. For this reason we will employ the corrected values to determine $a$. Using all available data on polarized lepton-nucleon scattering and applying the above mentioned corrections Ellis and Karliner obtain a best fit to the quark contribution to the nucleon helicity:

$$\Delta\Sigma \equiv g_A^0 = 0.37 \pm 0.07 \ . \tag{34}$$

Table 1 summarizes the experimental results on the various quantities, their relation with the parameter $a$ and the result for $a$ which is obtained from the respective values.

Fig. 2 shows the various values for the model parameter $a$. We also plot as a thick vertical line the mean from the results of the three independent experimental determinations, Eqs. 28, 29, and 34. The best known quantity is $g_A$. It is this value which therefore essentially determines the weighted mean for $a$:

$$a = 0.246 \pm 0.002 \ . \tag{35}$$

As one sees in Fig. 2 the various corrections are essential. By correcting the NMC result on $S_G$ for shadowing effect and the E142 result for $Q^2$-dependence the original values are shifted to within $1\sigma$ of the weighted mean of $a = 0.246$. We take this as evidence that the employed toy model captures at least the essential features of the idea of a pion cloud of the constituent quark and that this idea is capable of explaining some properties of the nucleon.

| Observable | Theory | Measurement | Result for $a$ |
|---|---|---|---|
| $S_G$ (NMC) | $S_G = (1/3)(1-2a)$ | $0.235 \pm 0.026$ | $0.148 \pm 0.039$ |
| $S_G + \delta^{\text{shadow}} S_G$ | | $0.154 \pm 0.026$ | $0.269 \pm 0.039$ |
| $g_A$ | $g_A = (5/3)(1-a)$ | $1.2573 \pm 0.0028$ | $0.246 \pm 0.002$ |
| $\Gamma_1^p$ (EMC) | $\Gamma_1^p = (5/18)(1-2a)$ | $0.126 \pm 0.025$ | $0.273 \pm 0.045$ |
| $\Gamma_1^n$ (SMC) | $\Gamma_1^n = -(5/18)a$ | $-0.08 \pm 0.08$ | $0.29 \pm 0.29$ |
| $\Gamma_1^n$ (E142) | | $-0.022 \pm 0.011$ | $0.079 \pm 0.040$ |
| $\Gamma_1^n$ (E142, corr.) | | $-0.056 \pm 0.011$ | $0.202 \pm 0.040$ |
| best fit for $g_A^0$ | $g_A^0 = 1 - 3a$ | $0.370 \pm 0.070$ | $0.210 \pm 0.023$ |

Table 1. Determination of $a$ by various observables.

However, it might come as a surprise that the probability for a $\pi^+$ in an up-constituent quark turns out to be almost 25 %. A perturbative calculation in the linear $\sigma$-model with pseudoscalar quark-pion-coupling (coupling constant $g_{\pi qq} = 3.76$) reproduces this value if one uses a regularization parameter of $\Lambda = 3.37$ GeV. This seems to be very large for a hadronic cut-off. A possible explanation for both large values might be the non-relativistic character of the employed models. It should be appropriate to describe the nucleon as consisting of three relativistic quarks. Relativistic quark models with light-cone wave functions for the constituent quarks have already been used for a quantitative analysis of static and quasi-static properties of the nucleon [21,22]. While such calculations can predict static properties of the nucleon, like $g_A$, almost exactly without recurring to an effective description like a pion cloud, they nevertheless cannot give a satisfactory agreement with experiment as far as quasi-static properties are concerned.

We conclude that even in a relativistic description of what constitutes a nucleon the idea of a pion cloud of the constituent quark might be helpful. However, the above performed estimates of the pion probability $a$ in the constituent quark, which are based on integral properties of the nucleon, are only of limited use. A direct and differential measurement of the pionic substructure of constituent quarks will offer more insight. To this purpose we turn in the next section to a differential description of the nucleon as seen in semi-inclusive electroproduction of pions on nucleon targets.

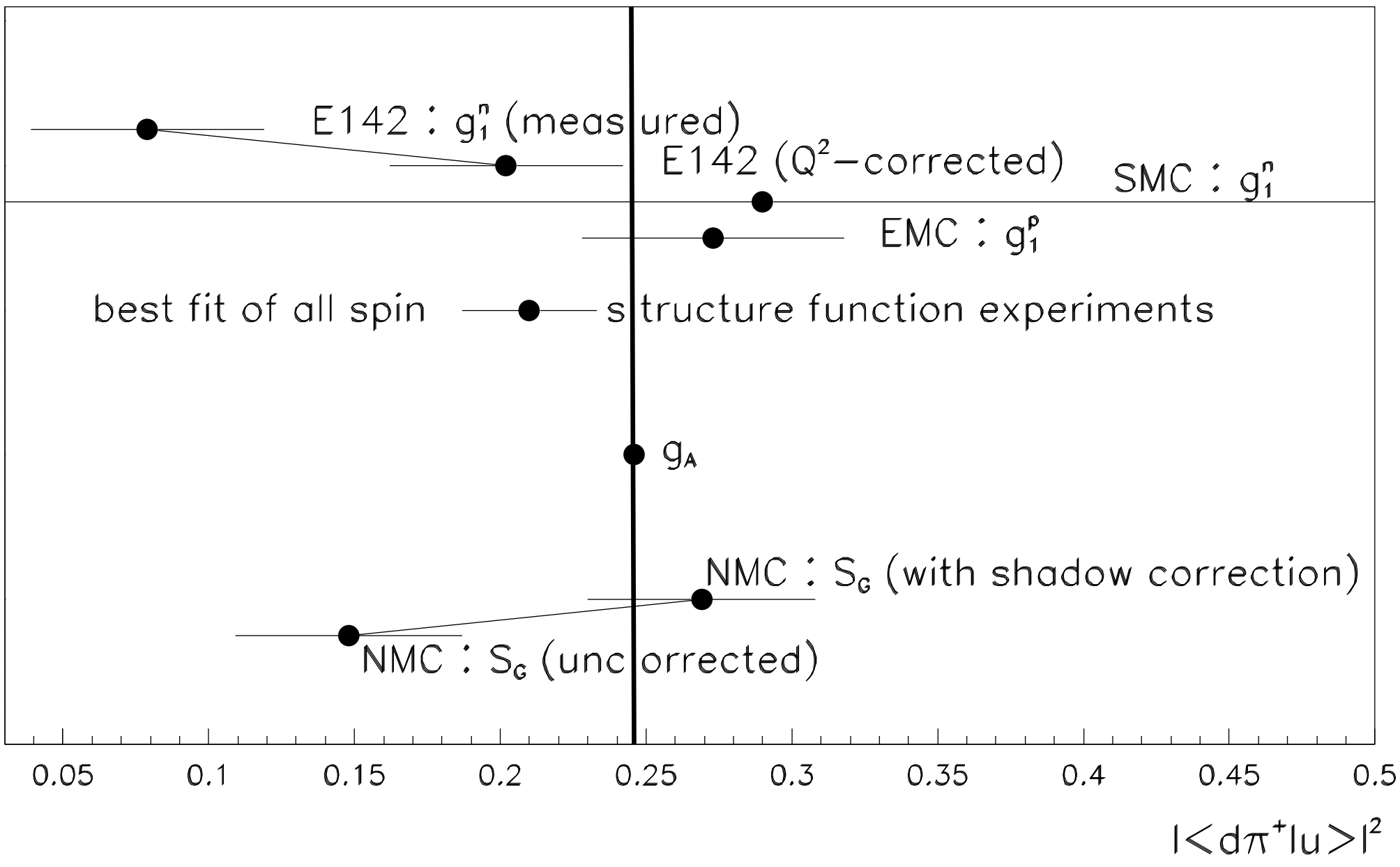

Figure 2. The various results for $a$ from different observables. The thick vertical line indicates $a = 0.246$.

## 3. Semi-Inclusive Pion Production

We consider electron scattering on the proton with at least one pion $\pi(k)$ in the final state:

$$\mathrm{e} + p \to e' + \pi + X \ . \tag{36}$$

Semi-inclusive pion production can be looked at as a soft hadronization process. Data [23] at high energies agree rather well with LUND-simulations. However, in certain kinematic regions higher twist effects can visibly modify the leading gluon cascade. For example, one expects contributions [24–26] from the hard fragmentation $\gamma^* q \to \pi q$.

We will describe the pion production in the naive parton model with the following assumptions. The nucleon consists of three valence constituent quarks, which are weakly bound. The square of their wave functions integrated over transverse momentum $\vec{p}_\perp$ up to the resolution $Q^2$ gives the probability to find a quark with flavor $q$ and light cone momentum fraction $x$ in the proton:

$$N_q(x) = \int_0^{Q^2} \mathrm{d}^2 p_\perp \, \varphi_q^*(x, \vec{p}_\perp) \, \varphi_q(x, \vec{p}_\perp) \ . \tag{37}$$

The constituents are massive Dirac particles ($M_Q = g_{\pi qq}\,\langle\sigma\rangle$) with a pointlike coupling to the photon and a pseudoscalar coupling $ig_{\pi qq}\,(\bar\psi\gamma_5\vec\tau\psi)\,\vec\pi$ to the pion. The coupling constant $g_{\pi qq}$ couples the quark equally to both fields of the $(\frac{1}{2},\frac{1}{2})$ $\mathrm{SU}(2)_L\times\mathrm{SU}(2)_R$ multiplet $(\sigma,\vec\pi)$. The coupling strength is determined from the empirical value of the constituent quark mass $M_Q = 350$ MeV and the value of $\langle\sigma\rangle = f_\pi = 93.3$ MeV:

$$g_{\pi qq} = 3.76\ . \tag{38}$$

Equally one could have estimated $g_{\pi qq}$ from quark additivity and the $\pi$-N-coupling constant $g_{\pi NN} = 13.5$:

$$g_{\pi qq} = \frac{g_{\pi NN}}{3\,g_A} = 3.57\ . \tag{39}$$

In the parton model, pion production is calculated as a folding of the partonic cross section $eq \to e'q'\pi$ with the distribution $N_q(x)$ of the quarks:

$$2E_\pi\,\frac{\mathrm{d}\sigma^{ep\to e'\pi X}}{\mathrm{d}^3k\,\mathrm{d}x_{\mathrm{B}}\,\mathrm{d}y} = \sum_q \int N_q(x)\,2E_\pi\,\frac{\mathrm{d}\sigma^{eq\to e'q'\pi}}{\mathrm{d}^3k\,\mathrm{d}x_{\mathrm{B}}\,\mathrm{d}y}\,\mathrm{d}x \tag{40}$$

The kinematical variables of this semi-inclusive process are on the electron side $x_{\mathrm{B}} = Q^2/2M_N\nu$ and $y = \nu/E$ and on the pion side $k^\mu = (E_\pi, \vec k)$. It is useful to discuss the kinematics in the next section in more detail.

### 3.1. Kinematics of Semi-Inclusive Pion-Production

In Fig. 3 we give a definition of all momenta in the target-at-rest system, $P^\mu = (M_N, \vec 0\,)$. In the semi-inclusive reaction there are two planes of interest, the lepton scattering plane $(\vec\ell, \vec\ell')$ and the photon-pion plane $(\vec q, \vec k)$. They include an angle $\phi_\pi$ between them. The energy of the incoming electron is $E$, the energy of the outgoing electron is $E'$. The photon energy is $\nu$, the pion energy is $E_\pi$. The pion momentum $\vec k$ is divided into a momentum parallel to the direction of the photon momentum $\vec q$

$$k_{||} = \frac{\vec k\cdot\vec q}{|\vec q\,|} \tag{41}$$

and transverse to the photon momentum $\vec q$

$$\vec k_\perp = \vec k - k_{||}\cdot\frac{\vec q}{|\vec q\,|}\ . \tag{42}$$

Unpolarized cross sections do not depend on the azimuthal angle of the scattered electron, therefore two invariants are sufficient to characterize the virtual photon. These are

$$\begin{aligned} Q^2 = -q^2 &= 4\,EE'\,\sin^2\theta_e/2 \\ \nu &= \frac{P\cdot q}{M_N} = E - E' \end{aligned} \tag{43}$$

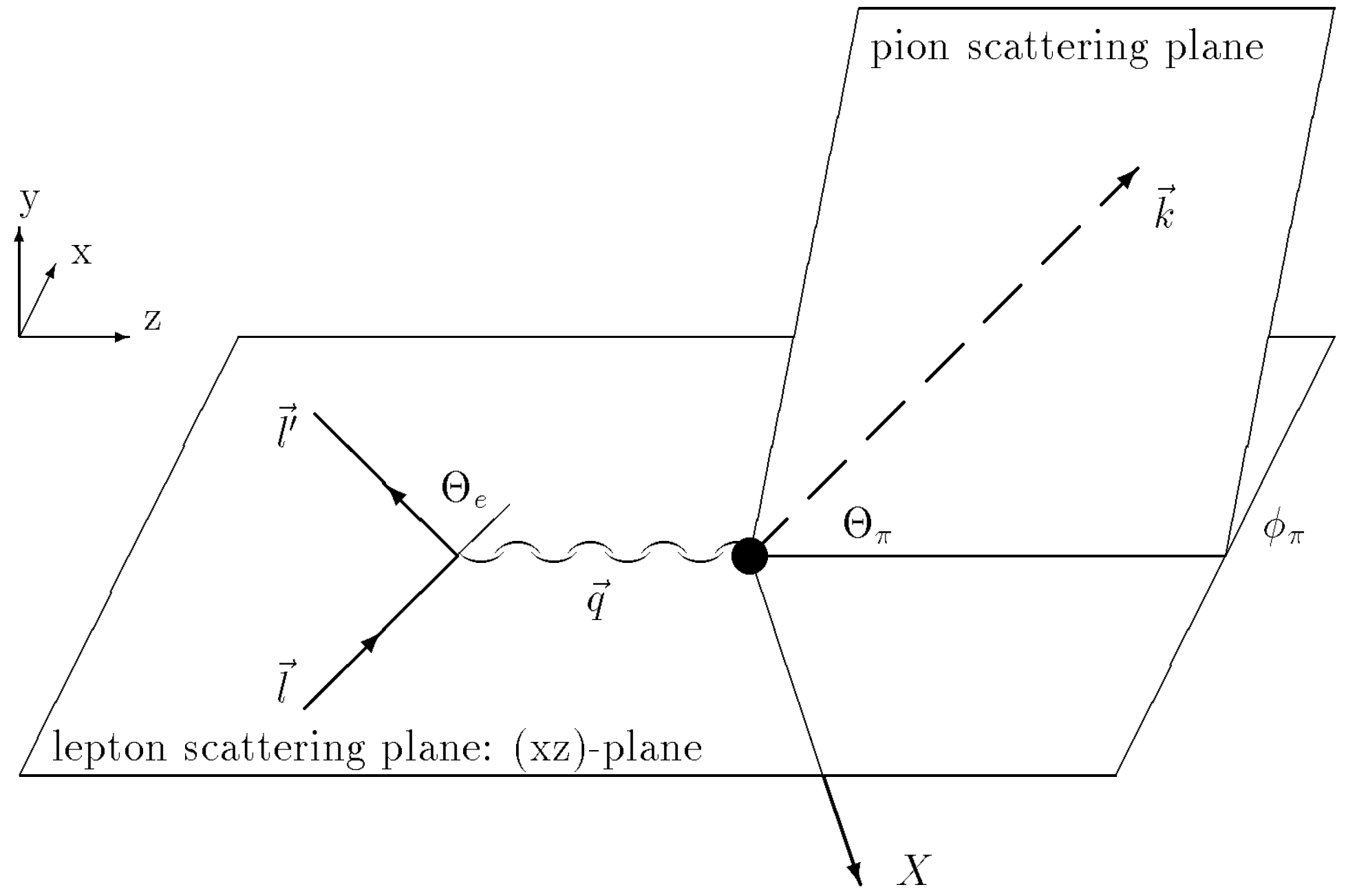

Figure 3. The reaction $ep \to e'\pi X$ in the target-rest- system.

or

$$x_{\mathrm{B}} = \frac{-q^2}{2P\cdot q} = \frac{Q^2}{2M_N\,\nu}$$
$$y = \frac{P\cdot q}{P\cdot \ell} = \frac{\nu}{E}\,. \tag{44}$$

The momentum of the pion has three components. We separate the azimuthal angle $\phi_\pi$ from the two residual variables $k_{||}$ and $|\vec{k}_\perp|$. The parallel momentum $k_{||}$ can be replaced by the energy variable $E_\pi$ or the scaling variable

$$z = \frac{P\cdot k}{P\cdot q} = \frac{E_\pi}{\nu}\,. \tag{45}$$

A conventional set of variables thus consists of ($Q^2$, $x_{\mathrm{B}}$, $z$, $|\vec{k}_\perp|$, $\phi_\pi$).

Another invariant variable besides $Q^2$, $P\cdot q$, and $P\cdot k$ is

$$q\cdot k = \nu\left\{E_\pi - \sqrt{1+\frac{Q^2}{\nu^2}}\,\sqrt{E_\pi^2-(\vec{k}_\perp^2+m_\pi^2)}\right\}$$
$$= \frac{zQ^4}{4M_N^2x_B^2}\left\{1-\sqrt{1+\frac{4M_N^2x_{\mathrm{B}}^2}{Q^2}}\sqrt{1-\frac{(\vec{k}_\perp^2+m_\pi^2)\,4M_N^2x_B^2}{z^2Q^4}}\right\}\,. \tag{46}$$

It can be simplified for $\xi \equiv x_{\rm B}^2 M_N^2/Q^2 \ll 1$, $\zeta \equiv \vec{k}_\perp^2 + m_\pi^2/Q^2 \ll 1$, and $\xi\zeta \ll z^2$ to

$$q \cdot k = \frac{1}{2}\left[ M_N^2\, z\, x_B^2 \; - \; Q^2\, z \; + \; \frac{\vec{k}_\perp^2 + m_\pi^2}{z} \right] \,, \tag{47}$$

which shows that in the Bjorken limit, $Q^2 \to \infty$, $q \cdot k = -\frac{1}{2} Q^2 z$ is no longer an independent variable.

It is advantageous to introduce the Mandelstam variables of the parton subprocess $\gamma^*(q) + q(p) \to \pi(k) + q'(p')$ (with $q \cdot k$ from Eq. 46):

$$\begin{aligned}
\hat{s} &= (p+q)^2 = M_Q^2 + Q^2 \left( \frac{x}{x_{\rm B}} - 1 \right) \\
\hat{u} &= (q-p')^2 = M_Q^2 + Q^2 \left( 1 - \frac{x}{x_{\rm B}} \right) + 2q \cdot k \\
\hat{t} &= (q-k)^2 = -Q^2 + m_\pi^2 - 2q \cdot k \,.
\end{aligned} \tag{48}$$

Note that $p$ and $p'$ denote the quarks' 4-momenta, whereas $P$ and $P'$ are nucleon 4-momenta. The Mandelstam variables are related by $\hat{s} + \hat{u} + \hat{t} = 2M_Q^2 + m_\pi^2 - Q^2$.

Since the incoming photon is a virtual particle with $q^2 = -Q^2 < 0$, the kinematical boundaries for the variables $x_{\rm B}$, $z$, $Q^2$, and $\vec{k}_\perp^2$ are interesting. The condition, that the quark *after* the production is on-mass-shell,

$$p'^2 = (xP + q - k)^2 = M_Q^2 \,, \tag{49}$$

expresses the quark momentum fraction $x$ in the nucleon to

$$\begin{aligned}
x \; = \; x_{\rm B} \Bigg\{ & \frac{1}{1-z} - \frac{m_\pi^2}{Q^2\,(1-z)} + \\
& \frac{Q^2\, z}{2M_N^2\, x_B^2\,(1-z)} \left( 1 - \sqrt{1 + \frac{4M_N^2\, x_{\rm B}^2}{Q^2}} \sqrt{1 - \frac{(\vec{k}_\perp^2 + m_\pi^2)\, 4M_N^2 x_{\rm B}^2}{z^2\, Q^4}} \right) \Bigg\} \,.
\end{aligned} \tag{50}$$

From the condition $x \leq 1$ follows a maximum $x_{\rm B}$ for a given set $Q^2$, $z$, and $k_\perp^2$. In Fig. 4 we show the maximal possible value of $|\vec{k}_\perp|$ for fixed $Q^2 = 1$ GeV$^2$. One sees that the allowed region for $x_{\rm B}$ becomes maximal for intermediate $z$ values. The maximal $|\vec{k}_\perp|$ is obtained for small $x_{\rm B}$.

### 3.2. Parametrization of Semi-Inclusive Pion Production

The inelastic electron scattering cross section can be factorized into a leptonic and hadronic tensor multiplied by the photon propagator squared

$$2E_\pi \, \frac{{\rm d}\sigma}{{\rm d}^3k \, {\rm d}x_{\rm B} \, {\rm d}y} = \frac{2\pi\alpha^2 M_N y}{Q^4} \, L_{\mu\nu} W^{\mu\nu} \tag{51}$$

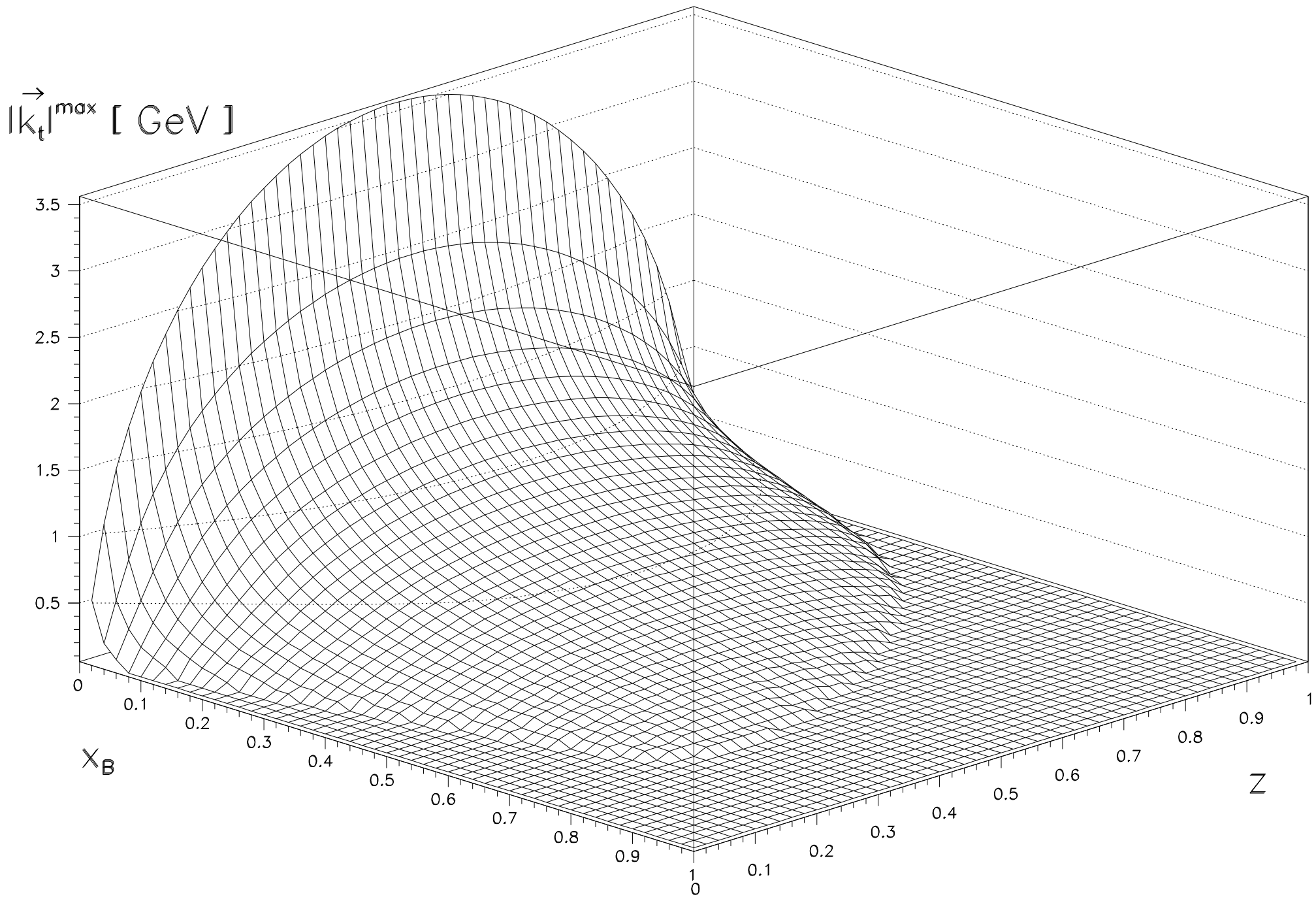

Figure 4. Maximal value of $|\vec{k}_\perp|$ as a function of $z$ and $x_{\rm B}$ at a fixed value of $Q^2 = 1\ {\rm GeV}^2$.

where

$$
\begin{aligned}
L_{\mu\nu} &= \frac{1}{2}{\rm Tr}\{\bar{u}(\ell')\gamma_\mu u(\ell)\,\bar{u}(\ell)\gamma_\nu u(\ell')\} \\
&= 2\ell_\mu\ell'_\nu + 2\ell_\nu\ell'_\mu - Q^2 g_{\mu\nu} \qquad (52)
\end{aligned}
$$

and

$$
\begin{aligned}
W^{\mu\nu} = \frac{1}{4M_N}\frac{1}{(2\pi)^4}\sum_{\sigma,X}\int \frac{{\rm d}^3P'}{(2\pi)^3\,2E_{P'}}\,&\langle X(P')\pi(k)|J^\mu_{em}(0)|N(P,\sigma)\rangle \\
\times\,\langle N(P,\sigma)|J^\nu_{em}(0)|X(P')\pi(k)\rangle \times (2\pi)^4\delta^4(P+q-k-P')\ . &\qquad (53)
\end{aligned}
$$

$W^{\mu\nu}$ is parametrized by four independent structure functions $W_i(x_{\rm B}, Q^2, z, \vec{k}^2_\perp)$ assuming parity and time reversal invariance [27]

$$
\begin{aligned}
W^{\mu\nu}(P,k,q) &= \left(\frac{q^\mu q^\nu}{q^2} - g^{\mu\nu}\right) W_1 + \frac{T^\mu T^\nu}{M_N^2} W_2 \\
&+ \frac{k^\mu_\perp T^\nu + T^\mu k^\nu_\perp}{m_\pi M_N} W_3 + \frac{k^\mu_\perp k^\nu_\perp}{m_\pi^2} W_4 \qquad (54)
\end{aligned}
$$

with the polarization vectors

$$
T^\mu = \left(P^\mu - \frac{P\cdot q}{q^2}q^\mu\right) \quad \text{and} \quad k^\mu_\perp = (0,\vec{k}_\perp,0)\ . \qquad (55)
$$

The cross section for definite photon polarizations can be obtained by contracting $W^{\mu\nu}$ with $\epsilon_\mu \epsilon_\nu$ of the wanted sort, e.g. $W_L = \epsilon_\mu(L)\, W^{\mu\nu} \epsilon_\nu(L)$ for the longitudinal photon polarization. Here are

$$
\begin{aligned}
\epsilon_\mu(L) &= (\nu, 0, 0, q_z)\, /\, \sqrt{Q^2} \\
\epsilon_\mu(x) &= (0, \vec{e}_x) \\
\epsilon_\mu(y) &= (0, \vec{e}_y)\ .
\end{aligned}
\tag{56}
$$

The longitudinal cross section and the purely transverse unpolarized $W_T = 0.5\,(\epsilon_\mu(x) W^{\mu\nu} \epsilon_\nu(x) + \epsilon_\mu(y) W^{\mu\nu} \epsilon_\nu(y))$ correspond to different combinations of $W_1$ and $W_4$, the interference term $W_{LT} \cos\phi_\pi = -\epsilon_\mu(x) W^{\mu\nu} \epsilon_\nu(L) + \epsilon_\mu(L) W^{\mu\nu} \epsilon_\nu(x)$ is proportional to $W_3$, and the transverse polarized term $W_{TT} \cos 2\phi_\pi = 0.5\,(\epsilon_\mu(x)\ W^{\mu\nu} \epsilon_\nu(x) - \epsilon_\mu(y) W^{\mu\nu} \epsilon_\nu(y))$ is proportional to $W_4$.

We transform the pion variables to $z$, $\vec{k}_\perp^2$, and $\phi_\pi$. With $\mathrm{d}^3k/2E_\pi = (\nu/4\,k_{||})\,\mathrm{d}z\,\mathrm{d}\vec{k}_\perp^2\,\mathrm{d}\phi_\pi$ the five-fold differential cross section has the form

$$
\begin{aligned}
\frac{\mathrm{d}\sigma}{\mathrm{d}x_{\mathrm{B}}\,\mathrm{d}y\,\mathrm{d}z\,\mathrm{d}\vec{k}_\perp^2\,\mathrm{d}\phi_\pi} &= \frac{4\pi\alpha^2 M_N E}{Q^4} \Big\{ x_{\mathrm{B}}\, y^2\, \mathcal{H}_1 + (1-y)\, \mathcal{H}_2 + \\
&\quad \frac{|\vec{k}_\perp|}{Q} (2-y) \sqrt{1-y}\ \cos\phi_\pi\, \mathcal{H}_3 + \\
&\quad \frac{\vec{k}_\perp^2}{Q^2} (1-y)\ \cos 2\phi_\pi\, \mathcal{H}_4 \Big\}
\end{aligned}
\tag{57}
$$

with

$$
\begin{aligned}
2z\,\mathcal{H}_1 &= M_N \left( W_1 + \frac{\vec{k}_\perp^2}{2m_\pi^2} W_4 \right) \\
2z\,\mathcal{H}_2 &= \nu \left( W_2 + \frac{\vec{k}_\perp^2}{2m_\pi^2} \frac{Q^2}{\vec{q}^{\,2}} W_4 \right) \\
2z\,\mathcal{H}_3 &= -\frac{Q^2}{m_\pi} W_3 \\
2z\,\mathcal{H}_4 &= M_N\, x_{\mathrm{B}}\, \frac{Q^2}{m_\pi^2} W_4\ .
\end{aligned}
\tag{58}
$$

In this paper we will concentrate on $\phi_\pi$ integrated cross sections which depend on $\mathcal{H}_1$ and $\mathcal{H}_2$ only. It is useful to define the combinations

$$
\begin{aligned}
\mathcal{H}_T &= \mathcal{H}_1 \qquad \text{and} \\
\mathcal{H}_L &= -\mathcal{H}_1 + \mathcal{H}_2/2x_{\mathrm{B}}\ ,
\end{aligned}
\tag{59}
$$

which correspond for inclusive scattering to $\mathcal{F}_T = \mathcal{F}_1$ and $\mathcal{F}_L = -\mathcal{F}_1 + \mathcal{F}_2/2x_{\mathrm{B}}$. To select the $\gamma$-$\pi$ interaction (see Fig. 5), one must analyze the longitudinal cross section. Since the pion is a boson and we neglect all transverse momenta, the longitudinal cross section will be given entirely by the $\gamma$-$\pi$ interaction.

### 3.3. Calculation of the Structure Functions $W_i(x_{\mathrm{B}}, Q^2, z, k_\perp^2)$

In analogy to Eq. 53 we define a partonic tensor

$$
\begin{aligned}
\hat{W}^{\mu\nu} &= \frac{1}{4x\,M_N}\,\frac{1}{(2\pi)^4}\sum_{\sigma\sigma'}\int\frac{\mathrm{d}^3p'}{(2\pi)^3\,2E_{p'}}\,\langle q(p',\sigma')\pi(k)|J^\mu_{em}(0)|q(p,\sigma)\rangle \\
&\quad\times\langle q(p,\sigma)|J^\nu_{em}(0)|q(p',\sigma')\pi(k)\rangle\times(2\pi)^4\delta^4(p+q-k-p') \\
&= \frac{1}{4x\,M_N}\,\frac{1}{(2\pi)^3}\sum_{\sigma,\sigma'}\langle q(p',\sigma')\pi(k)|J^\mu_{em}|q(p,\sigma)\rangle \\
&\quad\times\langle q(p,\sigma)|J^\nu_{em}(0)|q(p',\sigma')\pi(k)\rangle\;\delta(p'^2-M_Q^2) \\
&= \frac{1}{2x\,M_N}\,\frac{1}{(2\pi)^3}\,M^{\mu\nu}\,\delta(p'^2-M_Q^2)\;. \qquad (60)
\end{aligned}
$$

Here the last equation defines $M^{\mu\nu}$, which can be calculated from the Feynman diagrams of Fig. 5 with pseudoscalar coupling. The finite size of the pion is taken into account by the pion electromagnetic form factor $F_\pi(Q^2) = m_\rho^2/(Q^2+m_\rho^2)$.

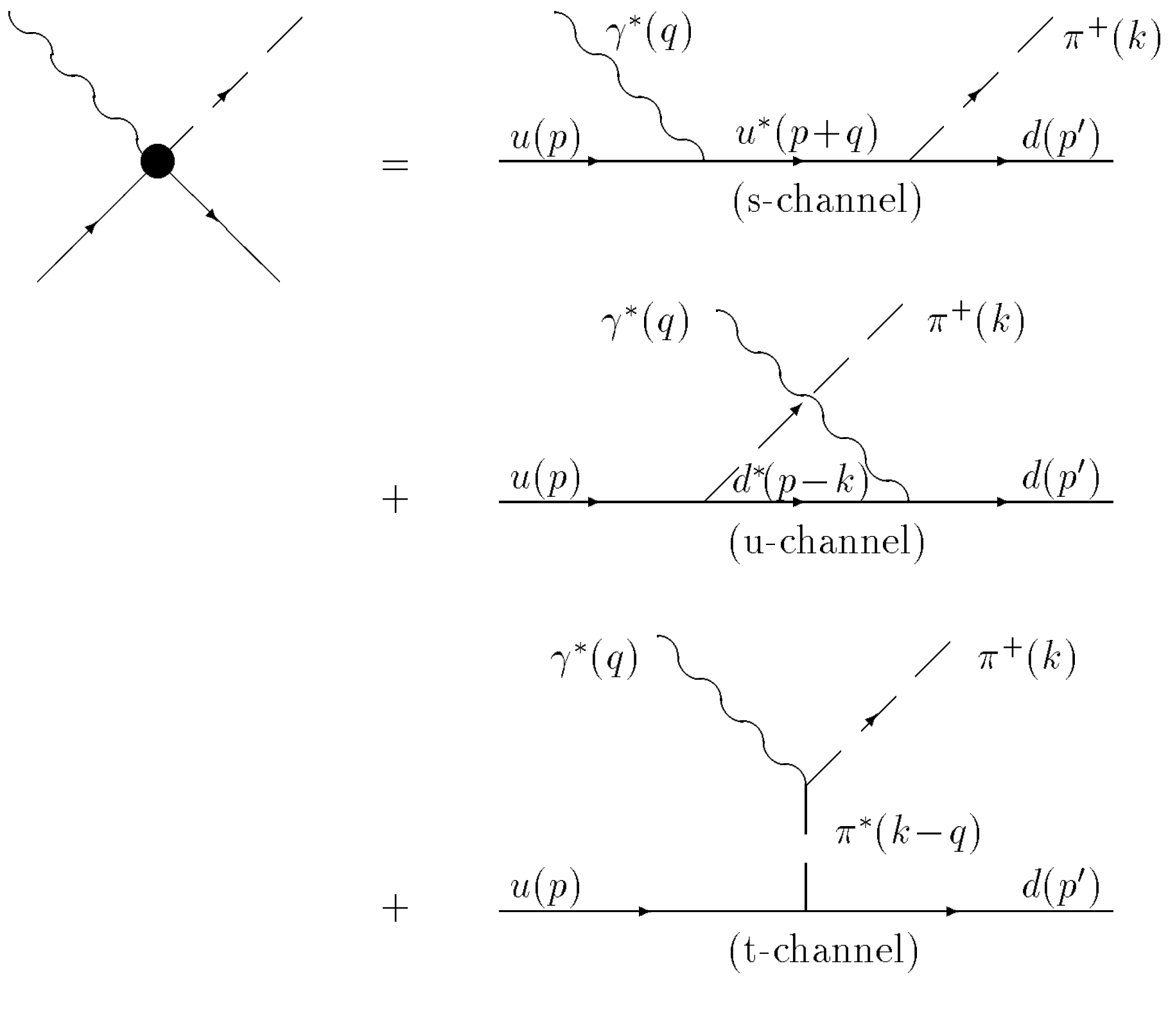

Figure 5. Tree level Feynman diagrams to the process $\gamma^* u \to d\pi^+$. Shown are the contributions from the s-channel, the u-channel, and the t-channel.

In addition, we suppress all off-shell propagators of intermediate state particles with monopole form factors:

$$
\begin{aligned}
g(p^2) &= g_{\pi qq}\,\frac{\Lambda^2}{\Lambda^2+(\hat{s}-M_Q^2)} \quad \text{in the s-channel} \\
&= g_{\pi qq}\,\frac{\Lambda^2}{\Lambda^2-(\hat{u}-M_Q^2)} \quad \text{in the u-channel} \\
&= g_{\pi qq}\,\frac{\Lambda^2}{\Lambda^2-(\hat{t}-m_\pi^2)} \quad \text{in the t-channel .}
\end{aligned}
$$

In general, the pion-quark-quark vertex depends on three momenta, $g(p^2, k^2, p'^2)$. In the three diagrams always two of the three particles are on-shell. We simplify the dependence on this off-shell momentum by parametrizing it with a monopole form factor containing one scale $\Lambda$. The role of $\Lambda$ has already been discussed in Section 2. The amount of a pion cloud in the u-quark depends logarithmically on $\Lambda$. We will use the value $\Lambda = 3.4$, which gives the best fit to the Gottfried sum rule and spin sum rule, see Section 2.

The deviation of the pion form factor $F_\pi$ from unity could lead to a violation of gauge invariance. One can, however, imagine adding a contact term of the form [28] $\sqrt{2}\, e\, g_{\pi qq}\, \epsilon_\pi\, \bar{u}(p', s')\, \gamma_5\, (F_\pi(Q^2) - 1)\, (q^\mu/Q^2)\, u(p, s)$ to restore gauge invariance. Since $\epsilon \cdot q = 0$, this term does not contribute to the calculation in Feynman gauge. A similar procedure has been used in Ref. [29].

For the process $\gamma^* u \to d\pi^+$ we calculate $M^{\mu\nu} = \frac{1}{2} Tr[M^\mu(\not{p} + M_Q)\, M^\nu\, (\not{p}' + M_Q)]$. With $q_u = 2/3$ and $q_d = -1/3$ as quark charges we get

$$
\begin{aligned}
M^\mu &= \sqrt{2}\, e\, g_{\pi qq} \left[ q_u\, \frac{\gamma_5(\not{p} + \not{q} + M_Q)\gamma^\mu}{\hat{s} - M_Q^2} + \right. \\
&+ \left. q_d\, \frac{\gamma^\mu(\not{p} - \not{k} + M_Q)\gamma_5}{\hat{u} - M_Q^2} + F_\pi(Q^2)\, \frac{2k^\mu \gamma_5}{\hat{t} - m_\pi^2} \right] . \qquad (61)
\end{aligned}
$$

The resulting tensor $M^{\mu\nu}$ can be reduced according to its Lorentz structure into four structure functions $\hat{W}_1$, $\hat{W}_2$, $\hat{W}_3$, $\hat{W}_4$ in the same way as pion production on the proton (cf. Eqs. 54,55). The equivalent polarization vector is $\hat{T}^\mu = x\,(P^\mu - q^\mu\, P \cdot q/q^2)$.

The structure functions $\hat{W}_i$ are listed in the Appendix. By folding the quark structure functions $\hat{W}_i$ with the quark probability distributions we obtain the hadronic structure functions $W_i(x, Q^2, z, \vec{k}_\perp^2)$

$$
\begin{aligned}
W_1^{p\to\pi^+ + X} &= \int_0^1 \mathrm{d}x\, \left[ N_u(x) \hat{W}_1^u + N_{\bar{d}}(x) \hat{W}_1^{\bar{d}} \right] \\
W_2^{p\to\pi^+ + X} &= \int_0^1 \mathrm{d}x\, x^2\, \frac{M_N^2}{M_Q^2}\, \left[ N_u(x) \hat{W}_2^u + N_{\bar{d}}(x) \hat{W}_2^{\bar{d}} \right]
\end{aligned}
$$

$$W_3^{p\to\pi^+ + X} = \int_0^1 \mathrm{d}x\, x\, \frac{M_N}{M_Q} \left[ N_u(x)\hat{W}_3^u + N_{\bar{d}}(x)\hat{W}_3^{\bar{d}} \right]$$
$$W_4^{p\to\pi^+ + X} = \int_0^1 \mathrm{d}x \left[ N_u(x)\hat{W}_4^u + N_{\bar{d}}(x)\hat{W}_4^{\bar{d}} \right] \tag{62}$$

These define the semi-inclusive structure functions (Eq. 58) and the differential cross section for $\pi$-production (Eq. 57).

## 4. The Cross Section for $\mathrm{ep} \to \mathrm{e}'\pi^+ X$

The five-fold differential cross section can be evaluated according to Eq. 57 with the four different structure functions $\mathcal{H}_i$. For a first comparison with future experimental data it will be necessary – for reasons of statistics – to reduce this five-dimensional space to the 3 dimensions $x_{\mathrm{B}}$, $Q^2$, and $z$ by integrating out $\phi_\pi$ and $\vec{k}_\perp^2$:

$$H_i\,(x_{\mathrm{B}}, Q^2, z) := \frac{1}{2} \int_0^{2\pi} \mathrm{d}\phi_\pi \int_0^{\vec{k}_\perp^2|_{\max}} \mathrm{d}\vec{k}_\perp^2\, \mathcal{H}_i(x_{\mathrm{B}}, Q^2, z, \vec{k}_\perp^2)$$
$$= \pi \int_0^{\vec{k}_\perp^2|_{\max}} \mathrm{d}\vec{k}_\perp^2\, \mathcal{H}_i(x_{\mathrm{B}}, Q^2, z, \vec{k}_\perp^2)\ , \tag{63}$$

where $\vec{k}_\perp^2|_{\max}$ is obtained from Eq. 50. The cross section is then given by

$$\frac{\mathrm{d}\sigma}{\mathrm{d}x_{\mathrm{B}}\,\mathrm{d}y\,\mathrm{d}z} = \frac{8\pi\alpha^2 M_{\mathrm{N}} E}{Q^4} \left\{ x_{\mathrm{B}}\, y^2\, H_1 + (1-y)\, H_2 \right\} \quad \text{or}$$
$$\frac{\mathrm{d}\sigma}{\mathrm{d}x_{\mathrm{B}}\,\mathrm{d}Q^2\,\mathrm{d}z} = \frac{4\pi\alpha^2}{x_{\mathrm{B}}\, Q^4} \left\{ x_{\mathrm{B}}\, y^2\, H_1 + (1-y)\, H_2 \right\}\ . \tag{64}$$

In Fig. 6 we show the two different contributions from $H_1$ and $H_2$ to the differential cross section of Eq. 64 as a function of $z$. As parameters we choose $E = 27$ GeV, $x_{\mathrm{B}} = 0.10$, and $Q^2 = 1$ GeV$^2$. The latter value is chosen as a compromise: $Q^2$ has to be minimal in order *not* to resolve the internal structure of the pion. The detection of the scattered electron, however, places a lower bound on $Q^2$. The value of $x_{\mathrm{B}}$ will become clear below.

The contribution to the cross section from $H_1$ is significantly lower than that from $H_2$. This is due to the kinematic factor $x_{\mathrm{B}}\, y^2$, which for our kinematic conditions is only about 0.004. Furthermore we note that the pion cloud in $t$-channel production contributes significantly to $H_2$, especially for large values of $z$. As discussed at the end of Section 3, this contribution is purely longitudinal owing to the pseudoscalar nature of the pion. This is obvious from Fig. 6, which shows a negligible contribution to $H_1 = H_T$.

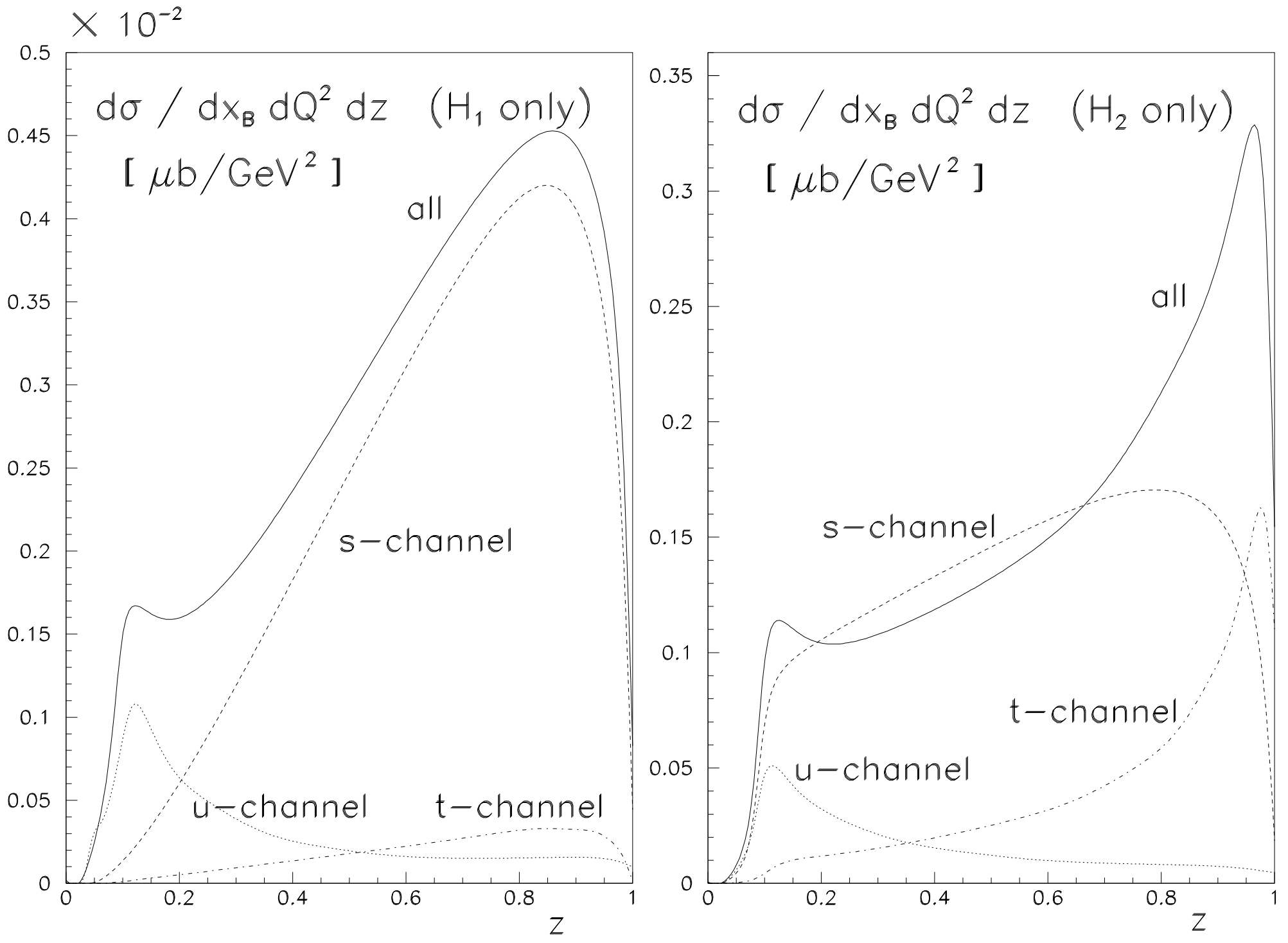

Figure 6. Contribution from $H_1$ and $H_2$ to the cross section for semi-inclusive pion production. The values of $Q^2$ and $x_{\rm B}$ are fixed at 1 GeV$^2$ and 0.10, respectively.

Of further interest is obviously the contribution to the cross section from longitudinal and transverse photon polarization, respectively. Decomposing the cross section into longitudinal and transverse structure functions we obtain:

$$\frac{\mathrm{d}\sigma}{\mathrm{d}x_{\rm B}\,\mathrm{d}Q^2\,\mathrm{d}z} = \frac{4\pi\alpha^2}{Q^4}\Big\{2(1-y)\,H_L + (1+(1-y)^2)\,H_T\Big\} \ . \tag{65}$$

The two contributions from $H_L$ and $H_T$ are plotted in Fig. 7. Clearly the contribution from $H_L$ at $z \to 1$ is dominated by the $t$-channel, whereas the $s$-channel contributes strongly to $H_T$. Thus a determination of the longitudinal part of the cross section and correspondingly the longitudinal structure function will allow to clearly isolate the presence and strength of a pion cloud. A recent calculation [25] of hadronization effects of the struck quark also showed evidence for a peak at large values of $z$ in the longitudinal structure function.

It should be stressed, that we require a large invariant mass of the hadronic system excluding the pion, such that the quasi-elastic channel $\gamma^* p \to \pi^+ n$ does not contribute. As an example, for $x_{\rm B} = 0.15$ at $E_e = 27\,{\rm GeV}$ this hadronic mass is only 1 GeV (2.4 GeV) for $z \to 1$ ($z \to 0$), with the hadronic mass increasing with decreasing $x_{\rm B}$. Thus a measurement at larger center-of-mass energy is desirable.

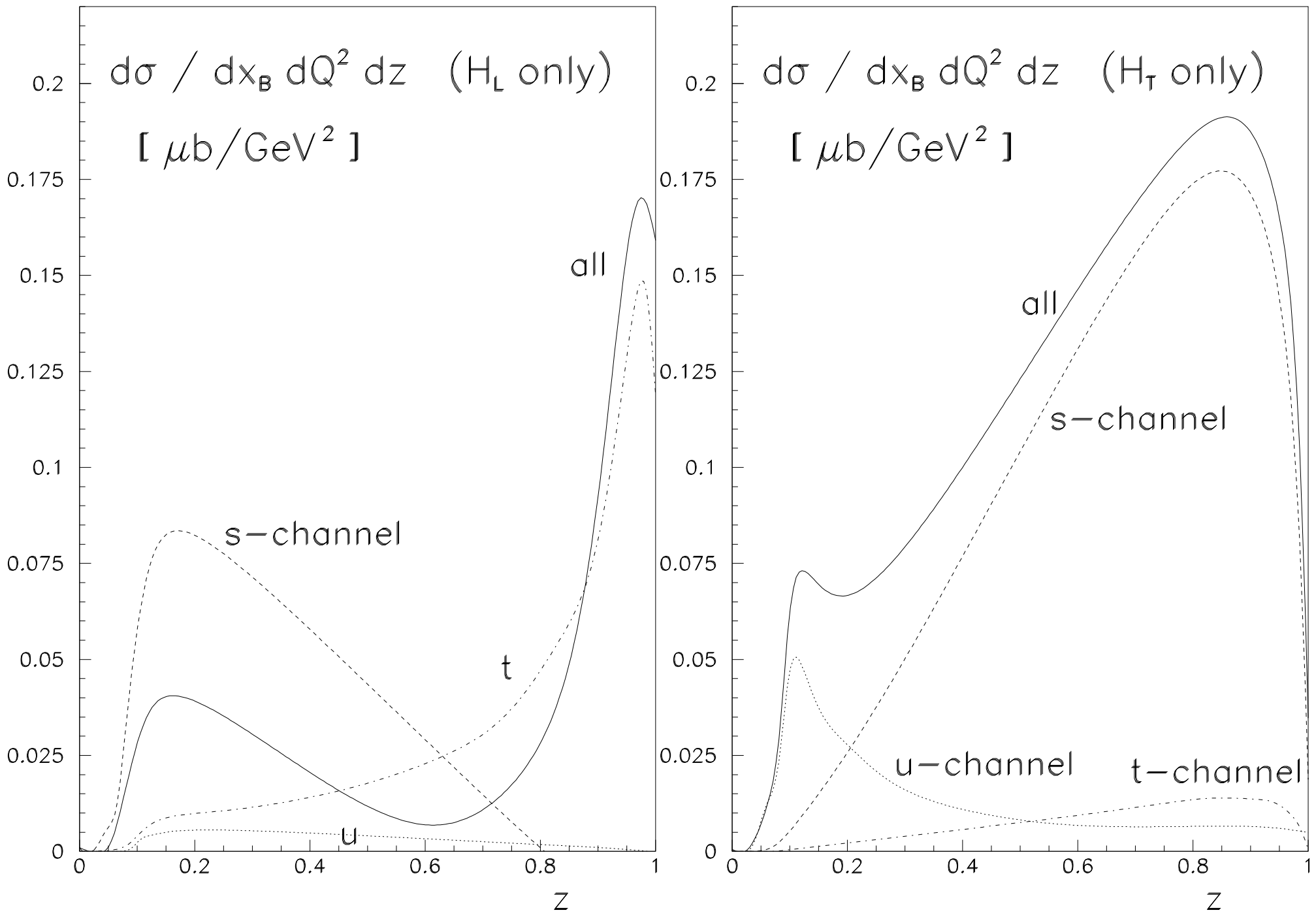

Figure 7. Contribution from $H_L$ and $H_T$ to the cross section for semi-inclusive pion production. The values of $Q^2$ and $x_{\rm B}$ are fixed at 1 GeV$^2$ and 0.10, respectively.

Finally the question arises which value of $x_{\rm B}$ is best suited to extract the signal from the underlying events. To this end we show in Fig. 8 the cross section as a function of $z$ for four different values of $x_{\rm B}$. Clearly the signal at large values of $z$ increases with decreasing $x_{\rm B}$. In addition, the cross section is largest for small values of $x_{\rm B}$. Thus it is desirable to perform this experiment at the smallest possible value of $x_{\rm B}$. Again, for a typical detector the smallest value is dictated by the smallest value of $Q^2$ accessible in the experiment. Using the HERMES detector as an example, the minimum $x_{\rm B}$ is about 0.03.

Furthermore, the value of $x_{\rm B}$ in the quark parton picture is related to $x_{\rm B} \simeq m(\text{parton})/M_N$, where the structure function $F_2$ has its maximum. Since in our case the parton is a pion we should choose $x_{\rm B} \simeq 0.15$. As a compromise we use $x_{\rm B} = 0.10$.

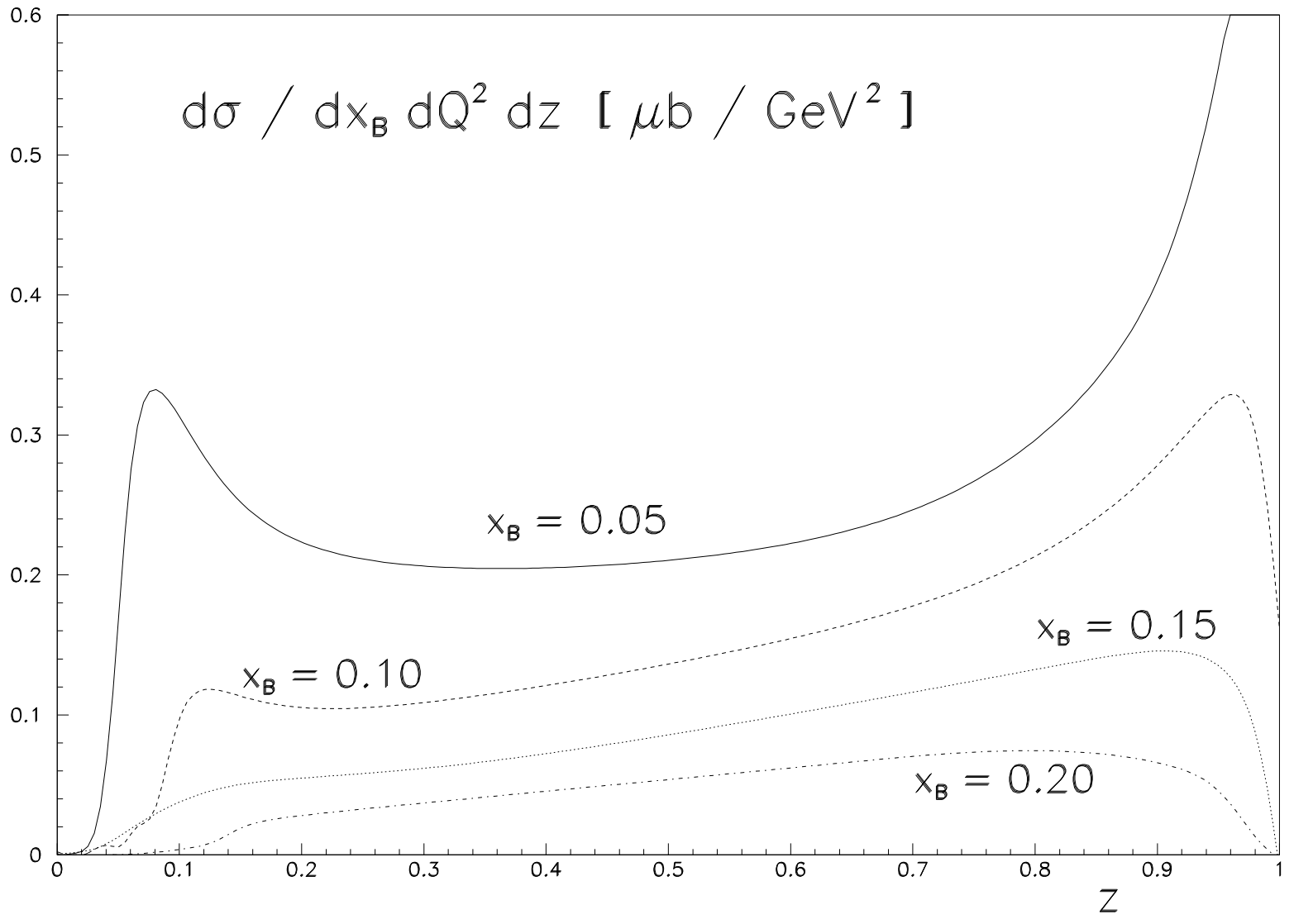

Figure 8. Cross section for semi-inclusive pion production as a function of the parameter $x_{\mathrm{B}}$. The value of $Q^2$ is fixed at 1 GeV$^2$.

### 4.1. Comparison of our Model with the LEPTO Event Generator

Experimental data on semi-inclusive pion production off nucleons is rather scarce. One of the main reasons for this lack of data was the limited luminosity and limited hadron identification of older experiments [30]. It is the new generation of experiments like the HERMES experiment [31] at HERA which have the possibility to operate at a high intensity circulating electron beam, which, together with a storage cell for hydrogen (or higher $A$-atoms), results in luminosities of $\mathcal{L} \simeq 1 \times 10^{31}\,/\mathrm{cm}^2\,\mathrm{s}$ for a polarized target. In case of an unpolarized target the luminosity may be increased by another factor of 10. Such high luminosities, together with the HERMES detector's capability to measure with large efficiency hadrons produced in electron-proton collisions results in the possibility for a detailed study of semi-inclusive measurements like the one proposed in this paper.

In the following we will compare the predictions of our model for the pion cloud of the quark with the results of a simulation using the Monte Carlo program LEPTO [32], which generates deep inelastic lepton-nucleon scattering events and hadronizes the final excited state using the Lund string model. We will demonstrate that our model yields cross sections of the same order as those simulated with LEPTO. Thus it should be possible to extract from data the amount of the quark's pion cloud. Of particular help will be the (near) vanishing of the *longitudinal* structure function, a fact known from several measurements.

In Fig. 9 we compare the semi-inclusive cross sections obtained from our model (dashed line) and from a simulation using the fastest $\pi^+$ from the LEPTO program (solid line). It is evident that the best 'signal to noise' at high $z$ is for low values of $x_B$. This is particularly true for the pion contribution from the $t$-channel, as can be seen from a comparison of the shapes of the $z$-distributions in Fig. 9 with those in Fig. 6. In addition it should be noted that the cross section is largest for small values of $x_B$.

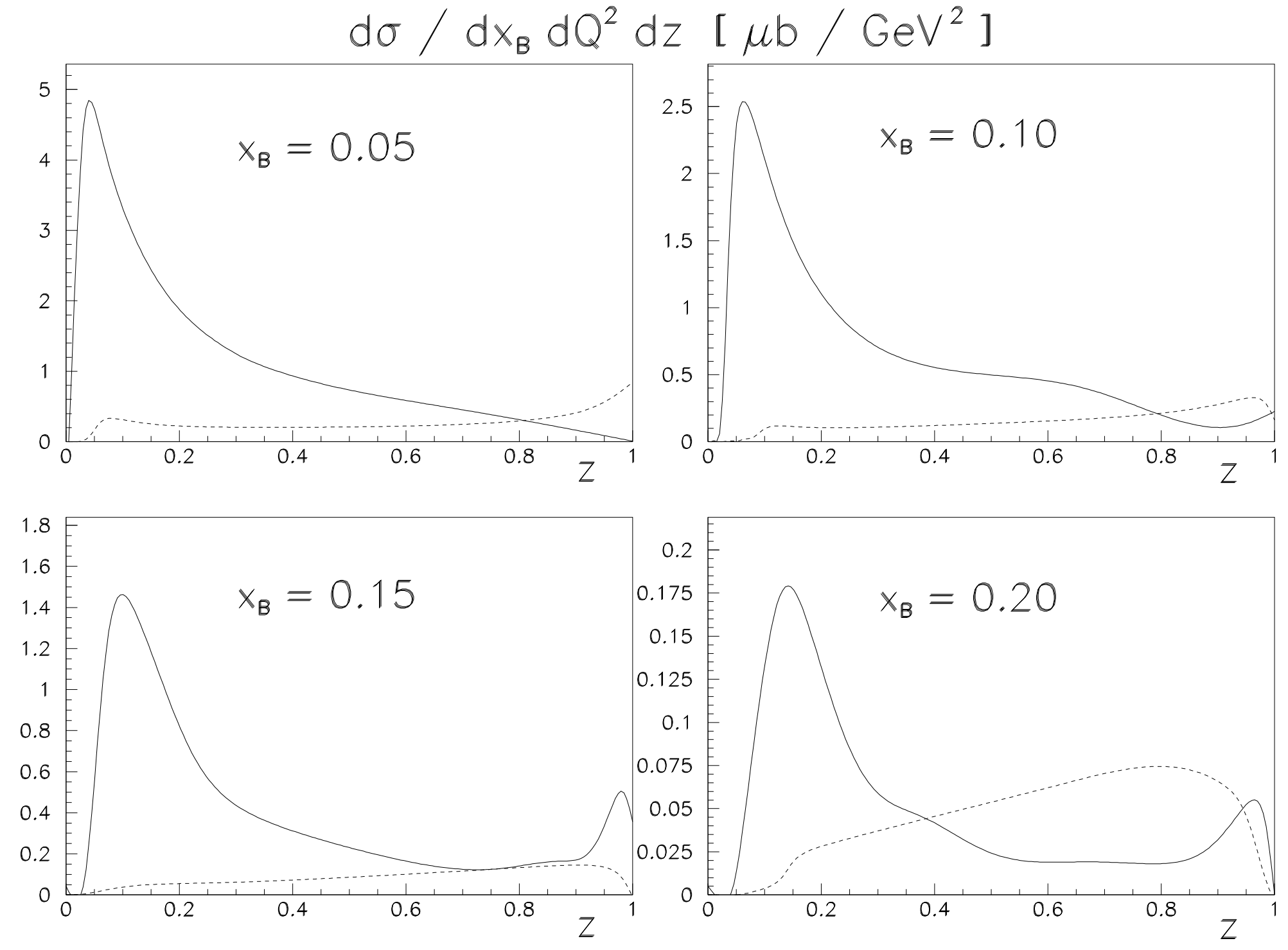

Figure 9. Comparison of semi-inclusive cross sections obtained from our model (dashed line) and from a simulation using the fastest $\pi^+$ from the LEPTO program (solid line). The value of $Q^2$ is fixed at 1 GeV$^2$.

Also note that the LEPTO event-generator produces the expected peak at large $z$ from the elastic channel $\gamma^* p \to \pi^+ n$ for a value of $x_B = 0.15$, corresponding to $x_B \simeq m_\pi / M_N$. In our model the wide s-channel 'continuum' at large $x_B$ values is equivalent to the hard fragmentation process, where one quark radiates a hard gluon which splits into a quark-antiquark pair; the antiquark then combines with the quark to form a pion [25]. Since for large $x_B$ values the hadronic mass excluding the pion is not large compared to the nucleon mass, the calculated cross section represents an *average* cross section over the produced resonances.

To get a feeling for the statistical accuracy of data, which can be accumulated in one month of data taking with an efficiency of 30% and a luminosity of $\mathcal{L} = 1 \times 10^{31}\,/\mathrm{cm}^2\,\mathrm{s}$, corresponding to $\int \mathcal{L} dt = 8 \times 10^6\ \mu\mathrm{b}^{-1}$, we show in Fig. 10 the expected $z$-spectra based on the LEPTO program and on our model. The data are integrated over $Q^2$ from 1 to 2 GeV$^2$ and over $x_{\mathrm{B}}$ from 0.05 to 0.15. A total of 270k events can be expected from LEPTO, whereas the pion model predicts 75k events. Clearly the statistical error, which is barely visible in Fig. 10, will allow a detailed shape comparison between data and LEPTO Monte Carlo, in order to ascertain a possible contribution from the pion cloud of the quark.

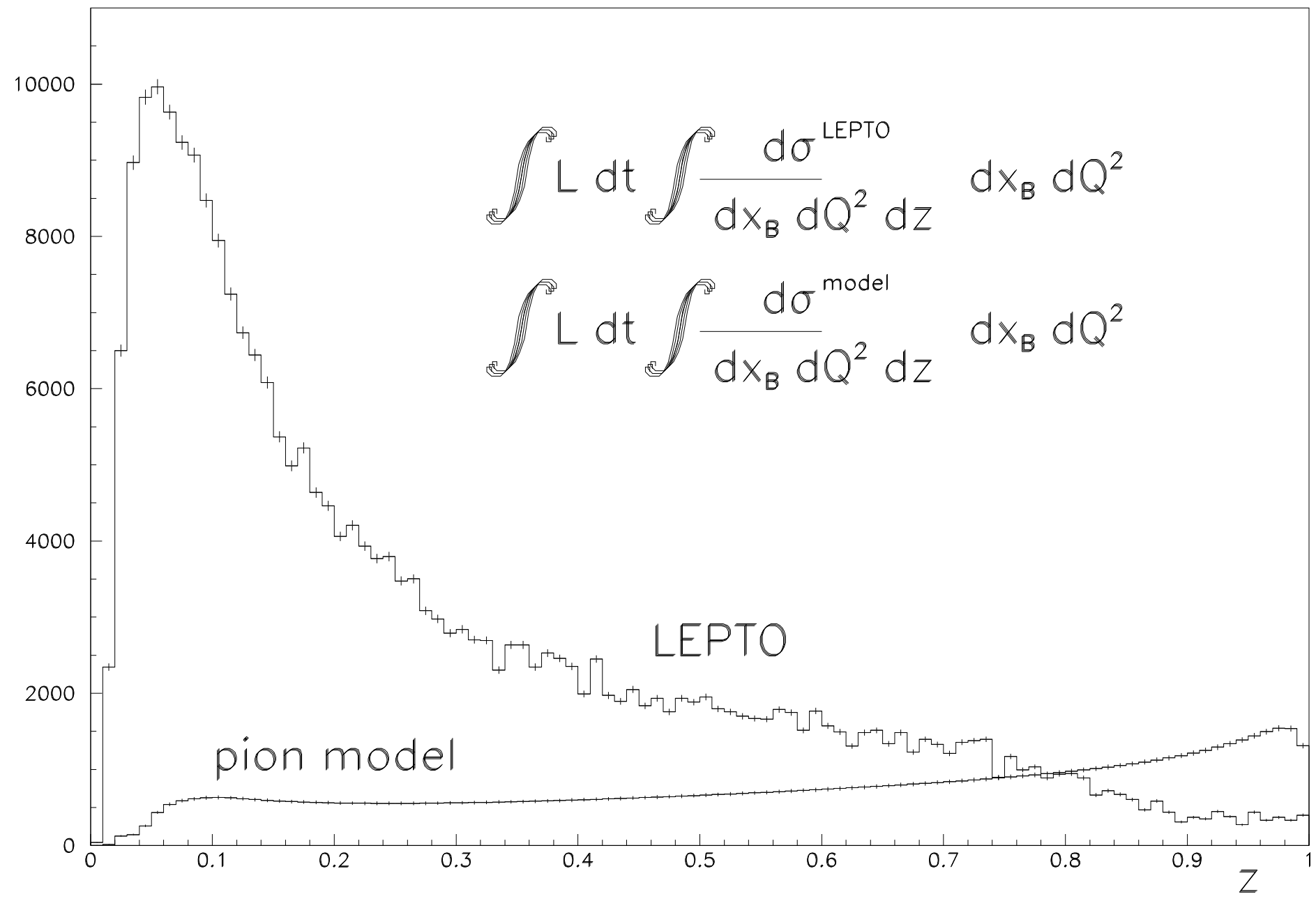

Figure 10. The expected $z$-spectra based on the LEPTO program and on our model. An integrated luminosity of $\int \mathcal{L} dt = 8 \times 10^6\ \mu\mathrm{b}^{-1}$ is assumed. The data are integrated over $Q^2$ from 1 to 2 GeV$^2$ and over $x_{\mathrm{B}}$ from 0.05 to 0.15.

However, the LEPTO Monte Carlo may not even describe the low $z$ distribution of events, which we believe are dominated by normal fragmentation events. In this case any extrapolation to large $z$ and any conclusion on a pion cloud contribution will be at least doubtful. In this case it will be necessary to separate from data directly the *longitudinal* structure function. This requires, however, additional data taking at about 15 GeV, again with an integrated luminosity of $\int \mathcal{L}dt = 8 \times 10^6\ \mu b^{-1}$. A preliminary analysis [33] shows, that with two such data sets, one at 27 GeV, the other at 15 GeV, a determination of the longitudinal structure function will be possible with an accuracy of about 20% in each of 100 bins of $z$. This will be more than sufficient in order to be able to extract the prominent feature of our model for the pion cloud of the constituent quark: a strong peaking near $z \to 1$ of the longitudinal structure function.

## 5. Conclusion

Employing a model where the physical quark fluctuates with some probability to a quark plus a pion we have calculated the differential cross section for semi-inclusive pion production in electron proton reactions. As expected, the biggest contribution from the pion cloud to the cross section is at small values of $x_B \simeq 0.1$ and large values of $z = E_\pi/\nu \to 1$. In particular the sole contribution to the *longitudinal* structure function will allow an experimental determination of the magnitude of the pion cloud of the quark.

## 6. Appendix

In the following we list the four independent structure functions $\hat{W}_i$ for the process $\gamma^* u \to d\pi^+$:

$$
\begin{aligned}
\hat{W}_1^u &= -\frac{g_{\pi qq}^2}{(2\pi)^3}\frac{1}{xM_{\rm N}}\left(\frac{q_u}{\hat{\hat{s}}}+\frac{q_d}{\hat{\hat{u}}}\right)^2\left(\hat{\hat{s}}\hat{\hat{u}}+m_\pi^2Q^2\right)\,\delta(p'^2-M_{\rm Q}^2) \\
\hat{W}_2^u &= \frac{4g_{\pi qq}^2}{(2\pi)^3}\frac{M_{\rm Q}^2}{xM_{\rm N}}\Bigg\{-m_\pi^2\left[\left(\frac{q_u}{\hat{\hat{s}}}+\frac{q_d}{\hat{\hat{u}}}\right)+\hat{\beta}\left(\frac{F_\pi}{\hat{\hat{t}}}-\frac{q_d}{\hat{\hat{u}}}\right)\right]^2 \\
&\qquad -Q^2\hat{\beta}\left[(1-\hat{\beta})\frac{q_uF_\pi}{\hat{\hat{s}}\hat{\hat{t}}}+\frac{q_d}{\hat{\hat{u}}}\left(\frac{F_\pi}{\hat{\hat{t}}}+\hat{\beta}\frac{q_u}{\hat{\hat{s}}}\right)\right] \\
&\qquad +\hat{\beta}(q_u-q_d-F_\pi)\left[\left(\frac{q_u}{\hat{\hat{s}}}+\frac{q_d}{\hat{\hat{u}}}\right)+\hat{\beta}\left(\frac{F_\pi}{\hat{\hat{t}}}-\frac{q_d}{\hat{\hat{u}}}\right)\right]\Bigg\} \\
&\quad\times\delta(p'^2-M_{\rm Q}^2) \\
\hat{W}_3^u &= \frac{2g_{\pi qq}^2}{(2\pi)^3}\frac{m_\pi M_{\rm Q}}{xM_{\rm N}}\Bigg\{(q_u-q_d-F_\pi)\left[\left(\frac{q_u}{\hat{\hat{s}}}+\frac{q_d}{\hat{\hat{u}}}\right)+2\hat{\beta}\left(\frac{F_\pi}{\hat{\hat{t}}}-\frac{q_d}{\hat{\hat{u}}}\right)\right] \\
&\qquad -Q^2\left[(2\hat{\beta}-1)\frac{q_uF_\pi}{\hat{\hat{s}}\hat{\hat{t}}}+\frac{q_d}{\hat{\hat{u}}}\left(\frac{F_\pi}{\hat{\hat{t}}}+2\hat{\beta}\frac{q_u}{\hat{\hat{s}}}\right)\right] \\
&\qquad -2m_\pi^2\left(\frac{F_\pi}{\hat{\hat{t}}}-\frac{q_d}{\hat{\hat{u}}}\right)\left[\left(\frac{q_u}{\hat{\hat{s}}}+\frac{q_d}{\hat{\hat{u}}}\right)+\hat{\beta}\left(\frac{F_\pi}{\hat{\hat{t}}}-\frac{q_d}{\hat{\hat{u}}}\right)\right]\Bigg\} \\
&\quad\times\delta(p'^2-M_{\rm Q}^2) \\
\hat{W}_4^u &= \frac{4g_{\pi qq}^2}{(2\pi)^3}\frac{m_\pi^2}{xM_{\rm N}}\left(\frac{F_\pi}{\hat{\hat{t}}}-\frac{q_d}{\hat{\hat{u}}}\right)\Bigg[(q_u-q_d-F_\pi)-m_\pi^2\left(\frac{F_\pi}{\hat{\hat{t}}}-\frac{q_d}{\hat{\hat{u}}}\right) \\
&\qquad +Q^2\frac{q_u}{\hat{\hat{s}}}\Bigg]\,\delta(p'^2-M_{\rm Q}^2)\ .
\end{aligned}
$$

Next we list the four independent structure functions $\hat{W}_i$ for the process $\gamma^*\bar{d}\to\bar{u}\pi^+$:

$$
\begin{aligned}
\hat{W}_1^{\bar{d}} &= -\frac{g_{\pi qq}^2}{(2\pi)^3}\frac{1}{xM_{\rm N}}\left(\frac{q_d}{\hat{\hat{s}}}+\frac{q_u}{\hat{\hat{u}}}\right)^2\left(\hat{\hat{s}}\hat{\hat{u}}+m_\pi^2Q^2\right)\,\delta(p'^2-M_{\rm Q}^2) \\
\hat{W}_2^{\bar{d}} &= \frac{4g_{\pi qq}^2}{(2\pi)^3}\frac{M_{\rm Q}^2}{xM_{\rm N}}\Bigg\{-m_\pi^2\left[-\left(\frac{q_d}{\hat{\hat{s}}}+\frac{q_u}{\hat{\hat{u}}}\right)+\hat{\beta}\left(\frac{F_\pi}{\hat{\hat{t}}}+\frac{q_u}{\hat{\hat{u}}}\right)\right]^2 \\
&\qquad +Q^2\hat{\beta}\left[(1-\hat{\beta})\frac{q_dF_\pi}{\hat{\hat{s}}\hat{\hat{t}}}+\frac{q_u}{\hat{\hat{u}}}\left(\frac{F_\pi}{\hat{\hat{t}}}-\hat{\beta}\frac{q_u}{\hat{\hat{s}}}\right)\right] \\
&\qquad +\hat{\beta}(q_u-q_d-F_\pi)\left[-\left(\frac{q_d}{\hat{\hat{s}}}+\frac{q_u}{\hat{\hat{u}}}\right)+\hat{\beta}\left(\frac{F_\pi}{\hat{\hat{t}}}+\frac{q_u}{\hat{\hat{u}}}\right)\right]\Bigg\} \\
&\quad\times\delta(p'^2-M_{\rm Q}^2) \\
\hat{W}_3^{\bar{d}} &= \frac{2g_{\pi qq}^2}{(2\pi)^3}\frac{m_\pi M_{\rm Q}}{xM_{\rm N}}\Bigg\{(q_u-q_d-F_\pi)\left[-\left(\frac{q_d}{\hat{\hat{s}}}+\frac{q_u}{\hat{\hat{u}}}\right)+2\hat{\beta}\left(\frac{F_\pi}{\hat{\hat{t}}}+\frac{q_u}{\hat{\hat{u}}}\right)\right] \\
&\qquad +Q^2\left[(2\hat{\beta}-1)\frac{q_dF_\pi}{\hat{\hat{s}}\hat{\hat{t}}}+\frac{q_u}{\hat{\hat{u}}}\left(\frac{F_\pi}{\hat{\hat{t}}}-2\hat{\beta}\frac{q_d}{\hat{\hat{s}}}\right)\right]
\end{aligned}
$$

$$
-2m_\pi^2 \left(\frac{F_\pi}{\hat{\hat{t}}} + \frac{q_u}{\hat{\hat{u}}}\right) \left[-\left(\frac{q_d}{\hat{\hat{s}}} + \frac{q_u}{\hat{\hat{u}}}\right) + \hat{\beta}\left(\frac{F_\pi}{\hat{\hat{t}}} + \frac{q_u}{\hat{\hat{u}}}\right)\right]\Bigg\}
$$

$$
\times\, \delta(p'^2 - M_{\rm Q}^2)
$$

$$
\hat{W}_4^{\bar{d}} = \frac{4g_{\pi qq}^2}{(2\pi)^3}\frac{m_\pi^2}{xM_{\rm N}} \left(\frac{F_\pi}{\hat{\hat{t}}} + \frac{q_u}{\hat{\hat{u}}}\right) \left[(q_u - q_d - F_\pi) - m_\pi^2 \left(\frac{F_\pi}{\hat{\hat{t}}} + \frac{q_u}{\hat{\hat{u}}}\right) - Q^2 \frac{q_d}{\hat{\hat{s}}}\right]
$$

$$
\times\, \delta(p'^2 - M_{\rm Q}^2)
$$

with the modified Mandelstam variables

$$
\begin{aligned}
\hat{\hat{s}} &= \hat{s} - M_{\rm Q}^2 \\
\hat{\hat{u}} &= \hat{u} - M_{\rm Q}^2 \\
\hat{\hat{t}} &= \hat{t} - m_\pi^2 \qquad \text{satisfying} \\
\hat{\hat{s}} + \hat{\hat{u}} + \hat{\hat{t}} &= -Q^2 \\
\text{and with}\quad q_u &= 2/3 \\
q_d &= -1/3 \qquad \text{as quark charges.}
\end{aligned}
$$

The factor of proportionality $\hat{\beta}$ is given by

$$
\hat{\beta} = \sqrt{\frac{k^2Q^2 + (q\cdot k)^2 + \vec{k}_\perp^2 Q^2}{p^2Q^2 + (p\cdot q)^2}} = \frac{E_\pi|\vec{q}\,| - k_{||}\nu}{xM_{\rm N}|\vec{q}\,|}\,.
$$